\documentclass[letterpaper,twocolumn,10pt]{article}
\usepackage{usenix-2020-09}

\usepackage{subcaption}
\usepackage{diagbox} 

\usepackage[binary-units=true]{siunitx}
\usepackage{multirow}
\usepackage{xcolor}
\usepackage[toc,page]{appendix}
\usepackage{enumitem}
\setlist[itemize]{noitemsep, topsep=0pt, leftmargin=0.5cm}
\usepackage{comment}
\usepackage{graphicx}
\usepackage{amsmath}
\usepackage{pifont} 

\newcommand{\cmark}{\ding{51}} 
\newcommand{\xmark}{\ding{55}} 

\definecolor{dred}{rgb}{0.75, 0.00, 0.00}
\definecolor{dgreen}{rgb}{0.00, 0.5, 0.00}
\definecolor{ddgreen}{rgb}{0.00, 0.50, 0.00}
\definecolor{dpink}{rgb}{0.75, 0.0, 0.75}
\definecolor{dblack}{rgb}{0.00, 0.00, 0.00}
\definecolor{dblue}{rgb}{0.00, 0.00, 0.75}
\definecolor{gyell}{rgb}{0.5, 0.5, 0.0}
\definecolor{dbleudefrance}{rgb}{0.19, 0.55, 0.91}
\definecolor{darkgoldenrod}{rgb}{0.72, 0.53, 0.04}
\definecolor{magenta}{rgb}{160,0,255}

\begin{document}
\pagenumbering{arabic}

\title{On the Generalizability of Machine Learning-based Ransomware Detection in Block Storage}


\author{
{\rm Nicolas Reategui}\\
EPFL
\and
{\rm Roman Pletka}\\
IBM Research - Zurich\\
\and
{\rm Dionysios Diamantopoulos}\\
IBM Research - Zurich\\
}


\definecolor{todocolor}{RGB}{219, 48, 122}
\newcommand{\TODO}[1]{\emph{\textcolor{todocolor}{\textbf{TODO:} #1}}}
\maketitle

\begin{abstract}
Ransomware represents a pervasive threat, traditionally countered at the operating system, file-system, or network levels. However, these approaches often introduce significant overhead and remain susceptible to circumvention by attackers. 
Recent research activity started looking into the detection of ransomware by observing block IO operations.  However, this approach exhibits significant detection challenges.
Recognizing these limitations, our research pivots towards enabling robust ransomware detection in storage systems keeping in mind their limited computational resources available. 
To perform our studies, we propose a kernel-based framework capable of efficiently extracting and analyzing IO operations to identify ransomware activity. 
The framework can be adopted to storage systems using computational storage devices to improve security and fully hide detection overheads. 
Our method employs a refined set of computationally light features optimized for ML models to accurately discern malicious from benign activities.

Using this lightweight approach, we study a wide range of generalizability aspects and analyze the performance of these models across a large space of setups and configurations covering a wide range of realistic real-world scenarios. We reveal various trade-offs and provide strong arguments for the generalizability of storage-based detection of ransomware and show that our approach outperforms currently available ML-based ransomware detection in storage.
Empirical validation reveals that our decision tree-based models achieve remarkable effectiveness, evidenced by higher median F1 scores of up to \SI{12.8}{\percent}, lower false negative rates of up to \SI{10.9}{\percent} and particularly decreased false positive rates of up to \SI{17.1}{\percent} compared to existing storage-based detection approaches. 
\end{abstract}



\sloppy

\section{Introduction}

Despite significant efforts being made by the industry to protect against cyberattacks, ransomware continues to be one of the top attack types encountered today~\cite{Security_2022}. Ransomware aims at encrypting and often also exfiltrating the victim's data, later asking for a ransom payment using cryptocurrencies to decrypt it and not disclosing the data publicly~\cite{Beaman_Barkworth_Akande_Hakak_Khan_2021, oz2022survey}. Existing cybersecurity defenses use a vast range of detection measures at the OS level by employing static signature-based techniques~\cite{berrueta2019survey}, observing process behavior patterns, file system accesses, API calls, analyzing honey-pot files, or by monitoring network activity~\cite{mcintosh2021ransomware, Hafeez}, HTTP logs~\cite{ongun2023celest, Oprea2018} and accessed DNS domains~\cite{Antonakakis2011, Rahbarinia}. Other approaches use behavioral fingerprinting of kernel events to detect anomalous patterns~\cite{sanchez2021fingerprinting, huertas2023}. Recently, ransomware detection in storage by observing block IO operations has been proposed on simple setups~\cite{hirano2019machine,Gagulic_Zumtaugwald_Sahu_2023} and cloud environments~\cite{zhongyu}.

Especially, for OS-level defenses, ransomware started to introduce techniques to bypass the defenses: Signature-based evasion techniques use polymorphic and metamorphic code to change the ransomware code whenever a system is attacked~\cite{popli19:behav_analysis_rw}. Other techniques perform obfuscation or encryption of the malicious code~\cite{banescu2015, pechoux} or extensively check the environment to assess the context in which the malware runs to adapt the attack strategy and hence possibly evade OS-level detection~\cite{lindorfer2011}. Fileless techniques such as memory-based execution prevent any signature analysis and leave no trace in the filesytem~\cite{sudhakar2020, Afianian}. Furthermore, advances in ransomware development have enabled attackers to run encryption from within a virtual machine (VM), hiding executables and payload from the host and spawning processed within the virtualized environment allowing them to escape detection~\cite{Symantec_2021}.  In addition, most OS-level detection mechanisms introduce a significant overhead resulting in either delayed detection or a drop in overall performance.


Relying heavily on OS-level defenses exposes the system to a large range of evasion techniques. Hence combining multiple independent detection methods is essential because one or more of the methods used can be bypassed or fail to detect an attack. Therefore, ransomware detection at the storage level reveals itself as a highly valuable defense for increasing security. Unfortunately, there is little work done on ransomware detection on storage.
Despite the limited information available from block IO operations, Hirano et al.~\cite{hirano2019machine} introduced a pioneering approach using various machine learning (ML) algorithms. 
Their ML models show high classification accuracy when using a small set of well-defined workloads, particularly an F1 score of \SI{98}{\percent} when using a Windows VM with NTFS. Subsequent research using a KVM hypervisor shows similar performance for additional setups \cite{Gagulic_Zumtaugwald_Sahu_2023}. Although these studies demonstrate good accuracy for the limited number of setups studied, they do not show if the models trained solely on these traces are able to generalize to a large number of system configurations. 
 To assess whether this high accuracy holds in a wider range of realistic setups, we conducted a motivational experiment where we applied Hirano's approach to different filesystems and applications.
As shown in Table \ref{tab:hirano_fs}, the F1 score of a model trained on NTFS drops dramatically when it is evaluated on Linux traces using EXT4 and XFS filesystems. Similarly, testing on unseen benign workloads yields an unexpected increase in false positive rate (FPR) 
as shown in Table \ref{tab:hirano_workloads}.

\begin{table}[b]
    \centering
    \captionsetup[subtable]{position = below}
    \captionsetup[table]{position=top}
    \caption{Results from a model trained with Hirano's approach.}
    \vspace{-7pt}
    \begin{subtable}{0.45\linewidth}
       \centering
       \begin{tabular}{|c|c|}
        \hline
        File-system & F1 \SI{}{\percent}  \\
        \hline
        NTFS~\cite{hirano2019machine}  & 98 \\
        EXT4 & 74.4  \\
        XFS & 70.39 \\
        \hline
        \end{tabular}
        \caption{F1 score using an XGBoost classifier trained on NTFS and evaluated on Linux traces using EXT4 and XFS.}
         \label{tab:hirano_fs}
   \end{subtable}
      \hspace*{1em}
    \begin{subtable}{0.45\linewidth}
    \centering
    \begin{tabular}{|c|c|}
        \hline
        Workloads & FPR \SI{}{\percent} \\ 
        \hline
        BZ2 & 4.83  \\
        MySQL & 39.44 \\
        PostgreSQL & 93.33 \\
        \hline
    \end{tabular}
    \caption{FPR results from an XGBoost classifier evaluated on BZ2 compression and SQL workloads.}
    \label{tab:hirano_workloads}
   \end{subtable}%

   \vspace{-2pt}
\end{table}

As we can see, for workloads and configurations the model has not been trained on, the accuracy can vary significantly and induce substantial false positive rates. The effects on detectability when using different filesystem types, the long-term behavior of file systems as files are added, moved, or deleted (thereby affecting the volume state), workload changes, the behavior of virtual environments, and the usage of file-system level encryption remain unclear. The configurations that generalize well in such a model and those that don't have not been studied. Knowing which configurations do not generalize is extremely important as it helps to collect traces that cover these configurations for training a robust model. 
Shedding light on this problem and providing guidelines for selecting training setups is the main objective of this paper.

Specifically, the contributions of this paper include:

\begin{itemize}
    \item We present a computationally efficient scheme for real-time ransomware detection that extracts minimal information on IO operations and uses appropriate engineering of IO-derived ML features. 
    \item We perform an in-depth generalizability analysis of ML models on multiple setups including different volume utilization, multiple file system types, long-term data placement effects from block allocation strategies in file systems, device encryption, and virtualization environments.
    \item We quantify the efficiency and accuracy of our real-time ransomware detection pipeline using the KVM hypervisor on Linux with Windows and Linux ransomware strains, showing low median FNR rates of \SI{0.9}{\percent} and \SI{1.95}{\percent} on NTFS and XFS.
\end{itemize}

In the remainder of this paper, we delve into the comprehensive methodology underpinning our ransomware detection pipeline (Section~\ref{sec:methodology}), from security considerations and data collection to inference execution, followed by a detailed exploration of our model's basic generalizability across varying volume states and file systems (Section~\ref{sec:generalizability}). Subsequent sections dissect the intricate effects of copy-on-write mechanisms in VM images (Section~\ref{QCOW_effects}), challenges posed by device encryption (Section~\ref{device_encrytion}), and the implications of deploying our solution in real-world settings (Section~\ref{sec:real}). A review of related work then contextualizes our contributions within the broader research landscape (Section~\ref{sec:related}), culminating in conclusions that encapsulate our findings and their significance for advancing ransomware detection technologies (Section~\ref{sec:conclusion}).

\section{Methodology}
\label{sec:methodology}

In this section, we first discuss the challenges of storage-based ransomware detection and present our architecture as well as the threat model. We then describe the feature engineering steps and the experimental setup.
Finally, we introduce the benign and ransomware workloads to be studied.

\subsection{Challenges}

Ransomware detection in block storage using features extreacted from IO operations is particularly challenging as semantic information is not available. Zhu et al.~\cite{Zhu2024} leveraged the flash translation layer (FTL) to bridge the semantic gap. Their FTL was used to collect IO traces with additional filesystem information; these traces were then classified in a trusted execution environment. The use of manually designed heuristics which are inherently filesystem-specific, leave its applicability to other file systems untested.

Machine learning approaches are well suited for detecting complex patterns for which rule-based methods are not efficient. Hirano et al.~\cite{hirano2019machine} extracted IO information from a limited number of setups and derived offline a small set of features to train ML models; these features include average Shannon entropy, average throughput (bytes written/read), and the variance of logical block addresses (LBA) accesses.
Similarly, ransomware encryption leads to identifiable LBA access patterns as it produces an excess of write-after-read operations\cite{zhongyu}. Gagulic et al.~\cite{Gagulic_Zumtaugwald_Sahu_2023} extended the feature set by including more statistical metrics such as the kurtosis and slope of the distribution; these metrics better capture changing trends on entropy and LBA accesses within the time window. Nevertheless, the significant computational overhead prevents efficient real-time detection.

Despite the reported high performance of ransomware detection in storage  in these approaches, multiple effects - in particular the impact from different file systems or device states - remain to be elucidated. First, Linux and Windows file systems use different block allocation strategies. This can potentially impact the performance of a classifier tested on a file-system not included at training. 
Second, the long-term data placement effects from block allocation strategies play an important role in the overall data distribution across the available sectors. Over time, files are scattered, fragmented, or sparsely grouped across the device's LBA space which can impact LBA-derived features and reduce the model's accuracy.

Furthermore, keeping low FPR is essential for ransomware detection pipelines, as even a low FPR can lead to large number of alerts per day  
unmanageable when administrating a large fleet of storage systems. 
Therefore, our analysis focuses on improving the generalizabily of the first-level classifier.


\subsection{Architectural overview}

Our ransomware detection architectures a and b, illustrated in Figure~\ref{fig:arch_overview}, use a RHEL-based (Red Hat Enterprise Linux) server with our device-mapper kernel module \texttt{dm-entropy}; this module performs in-line entropy calculation on write IO operations and communicates with \texttt{SystemTap} for tracing and extracting information~\cite{prasad2005systemtap}. Further isolation can be obtained through virtualized environments running on top of a hypervisor (Figure~\ref{fig:arch_overview}(b)). 
We have implemented our framework in a commercially available storage systems with computational storage devices (CSD) providing an even more robust detection pipeline at no measurable overhead (Figure~\ref{fig:arch_overview}(c)). 
The CSDs enable in-line feature extraction directly in hardware with no impact on host IO operations. ML inference is performed by a dedicated storage controller within the storage system; this allows us to offload the detection overhead from the OS. Additionally, in storage systems, we can address potential challenges associated with a RAID setup by excluding all parity related operations during feature computation.

\subsection{Security Aspects}

\begin{figure}[t]
  \centering
  \includegraphics[width=\linewidth]{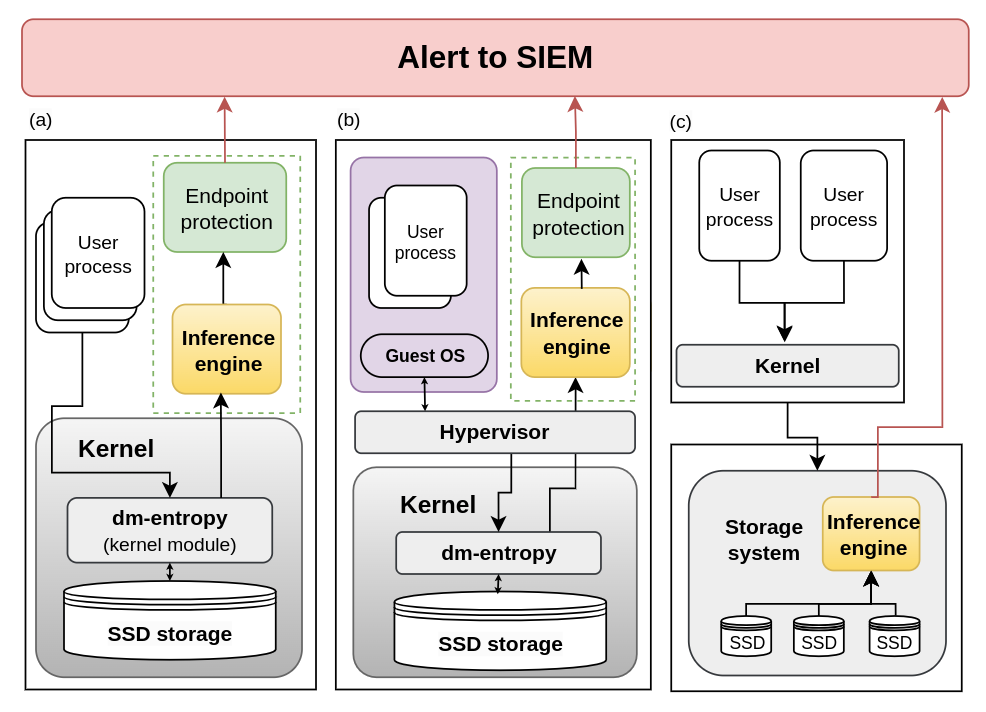 }
  \vspace{-20pt}
  \caption{(a) Linux-based ransomware detection in user space. (b) Hypervisor-based detection of ransomware on guest OS. (c) Storage system with integrated detection capabilities using CSD SSDs. }
  \vspace{-20pt}
  \label{fig:arch_overview}
\end{figure}

The proof-of-concept architecture (a) in Figure~\ref{fig:arch_overview} performs the feature extraction on IO operations in a kernel module. Successful detection is possible on compromised systems where adversaries have no root privileges. Figure~\ref{fig:arch_overview}(b) provides an additional isolation layer; our threat model assumes the guest OS can be fully compromised, potentially disabling defenses within the VM but considers the virtualization layer provided by QEMU/KVM as a trusted entity. The virtualization layer ensures the execution of our detection pipeline. Finally, our target architecture in Figure~\ref{fig:arch_overview}(c) provides strong isolation where the host system can be fully compromised. 
As the inference engine is fully integrated into the storage system, the interference and reporting are no longer executed on a potentially compromised system, and only block IO operations are passed to the storage system. We assume the storage system administrator is a trusted actor responsible for managing access to the system as well as reporting ransomware incidents and not running on the host system.
The ransomware detection of all architectures can be integrated into security information and event management tools~\cite{SIEMs, Oprea2018}. 
Furthermore, implementing ransomware-specific incident responses~\cite{Bajpai_Enbody_2023} to protect from ransomware attacks allows for a much-needed systematic incident response strategy for Storage as a service (StaaS). 

\subsection{Feature Engineering}


Our approach first extracts minimal information from each IO operation including entropy of writes, LBA, type of IO (i.e., read or write), transfer size as well as a time stamp. We then compute 79 lightweight features over a window within the seconds range. 
Entropy-related features account for 19 of these and include average, mean absolute deviation (MAD), and 16-bin histograms. We also calculate the mean entropy of rewrites computed over one epoch on LBAs seeing a read IO before the write. Similarly, we compute the mean and a 17-bin histogram for LBA values; in total we extract 18 features for each read and write. 
Finally, we compute 17 transfer size features for each IO type including the average and an 11-bin histogram. 
Our CSDs further extract information on data compressibility as it is provided at no cost.
We introduce histograms as they are computationally efficient and can provide more information about the underlying distribution. 
Using these highly expressive features is necessary when coarse metrics are no longer reliable; for example, Lockfile uses intermittent encryption which increases the speed and lowers the mean entropy values. Additionally, we include file-system information as one-hot encoding for NTFS, XFS, and EXT4; this results in a total of 82 features. As the file-system information is not provided by IO operations, we extract it by scanning the initial sectors of volumes. File-system awareness is used throughout the multiple analyses unless denoted specifically. 
\par 
Our set of features can be efficiently calculated inline in $O(n)$ complexity. 
For example, entropy histograms are calculated by counting the occurrences of all 256 possible values resulting in a complexity of O($n$), $n$ being the number of Bytes.
In our CSDs, information on each IO operation is extracted directly in hardware and feature information is then summarized in the background by two dedicated Cortex R5 processors from extracted information. Therefore, this approach has no impact on host IO performance.

The effects of varying window sizes between 1 and \SI{60}{\second} have been extensively studied and are well understood~\cite{hirano2022ransap, Gagulic_Zumtaugwald_Sahu_2023}: Large windows better approximate the distributions likely leading to higher accuracy but also leads to larger detection latency. 
On the other hand, a small window size enables faster detection. An ML model trained on feature vectors extracted over this window, also denoted as an epoch, can be augmented with additional post-processing to further improve the overall accuracy by using a second-level classifier~\cite{zhongyu}. 


We explored multiple decision tree-based architectures such as Random Forest, and Gradient Boosting Machines. 
Nevertheless, model selection within decision-tree algorithms has a considerably lower impact on performance compared to training on different system configurations. 
Prior work has shown that XGBoost outperforms Random Forest by \SI{0.9}{\percent} on average~\cite{Gagulic_Zumtaugwald_Sahu_2023}, which is consistent with our findings.
We therefore limit our analysis by using an XGBoost classifier as a boosting algorithm for a fair comparison with state-of-the-art and better readability. Additionally, we use k-fold validation ($k=5$) in all our experiments.
We have observed that hyper-parameter optimization for the XGBoost model improves F1 scores by up to \SI{2}{\percent}. We use these optimized parameters for all results presented in this analysis.
However, we expect that including a larger number of setups will increase the need for an extensive analysis of hyper-parameter optimization.

\subsection{Experimental setup}

To have a robust pipeline for storage-based ransomware detection, we assess the impact of different volume states (Section~\ref{volume_state}) and file systems (Section \ref{filesystem_generalizability}). We collect IO traces, on either XFS or EXT4, using Setup a from Figure \ref{fig:arch_overview}.
For Windows-based workloads, we use a Windows 10 VM on QEMU/KVM that corresponds to Setup b in Figure \ref{fig:arch_overview}.
Furthermore, our assessment with real ransomware samples is performed using isolated VMs for security reasons. 

We study long-term data placement effects through a simple approach in which we sequentially copy and remove files resulting in file-system specific data fragmentation. Although more sophisticated aging methods have been developed including \textit{Impressions}~\cite{impressions}, and more recently \textit{Geriatrix}~\cite{geriatrix} which also induces free space segmentation, these techniques are extremely time consuming. 
Our approach turns out to be sufficient to demonstrate aging effects from file-system data placement.
We also study the impact of disk utilization by copying half of the data set to a different location on the same device, increasing the disk utilization by \SI{48}{\percent} (from \SI{52}{\percent} to \SI{77}{\percent}) of a \SI{1}{\tera\byte} device. The additional data is not subjected to any workload and is meant to represent data regions not attacked by ransomware but contributing to the overall volume utilization.

We study the generalizability of our models on \textit{Online Transactional Processing} (OLTP) workloads (Section~\ref{workloads_generalizability}) and the impact of copy-on-write formats (Section~\ref{QCOW_effects}). VMs running on QEMU/KVM or other virtualized environments commonly store data using these formats. We collect traces using a \texttt{qcow2} Windows 10 VM where we executed benign and ransomware workloads. 
Finally, we assess the impact on detectability when performing device encryption - a common setup used by organizations dealing with sensitive data - using \texttt{LUKS} 
(Section ~\ref{LUKS_encrytion}) and \texttt{Bitlocker} (Section ~\ref{QCOW_bltk}).

All training and test data is balanced w.r.t. the number of ransomware and benign labels but also regarding file-system type as well as workload types unless denoted otherwise when studying the effect of variations on that specific modality. Also, specific sections of the traces are either used for training or testing.

\subsection{Benign and ransomware workloads}

The first class of benign type of workloads consists of automated file conversions from a large database. 
We use files from the Govdocs dataset~\cite{garfinkel:2009} which consists of around 1 million files including images, office documents, and text files in various formats. 
The goal is to simulate real-world activity by converting files having different extensions resulting in variable entropy writes and file-system specific data placement.
To cover both low and high-throughput workloads we used different concurrency levels.  Furthermore, we study file compression using different compression algorithms. These workloads can be challenging for ransomware detection pipelines based solely on entropy-related features as compression results in higher entropy. Finally, we wanted to include benign workloads resembling typical OLTP database systems. To do so we use the \texttt{sysbench} benchmark tool~\cite{sysbench} which has multiple configurable parameters allowing us to cover a large number of OLTP setups. We use \texttt{sysbench} with \texttt{MySQL} and \texttt{PostgreSQL} drivers and vary the number of threads, tables, and sizes as well as enable \texttt{MySQL} database compression.

For security reasons and to enhance scalability for automated IO trace collection, we decided to use WannaLaugh, a ransomware emulator capable of creating a variety of ransomware-like IO patterns~\cite{diamantopolous2024wannalaugh}. This emulator is highly configurable to closely mimic real ransomware by choosing from a wide range of options for directory selection and file ordering, encryption algorithm, and write methods allowing entire, partial, or intermittent file encryption. We can emulate IO patterns from real ransomware such as Black Basta, Conti, Lockbit, Lockfile, or WannaCry. We further collected traces manually for a range of ransomware samples as will be described in Section~\ref{real_ransomware_validation}.




\section{Basic generalizability aspects}
\label{sec:generalizability}

In this section, we train multiple ML models on specific system configurations and assess their performance across volume states, file-system types, and various workloads.
Although we are mainly interested in the performance of the model including all configurations, we train additional models by specifically excluding certain configurations. These models hence provide excellent insights on the generalizability w.r.t the configurations.
The comparison with Hirano's approach across all studied setups shows that ours leads to higher median F1 scores of up to \SI{12.8}{\percent}, lower FNR of up to \SI{10.9}{\percent} and more importantly decrease in FPR of up to \SI{17.1}{\percent}.
\vspace{-5pt}

\subsection{Volume state generalizability} \label{volume_state}


In a preliminary analysis comparing the file-system fragmentation of Windows and Linux we observe significant differences in the proportion of fragmented files on a freshly created volume. NTFS is characterized by a large fragmentation and a long-tailed distribution of fragments per file. On the other hand, EXT4 has considerably fewer fragments, and XFS has no fragmentation. 
Furthermore, our procedure simulating long-term data placement disproportionately affects NTFS; it significantly increases the fragments per file compared to EXT4 or XFS. As file-system fragmentation impacts LBA locations, it is key to analyze the performance of our models on all volume states.


\begin{figure}[!b]
    \vspace{-10pt}
	\centering
	\includegraphics[width=\linewidth]{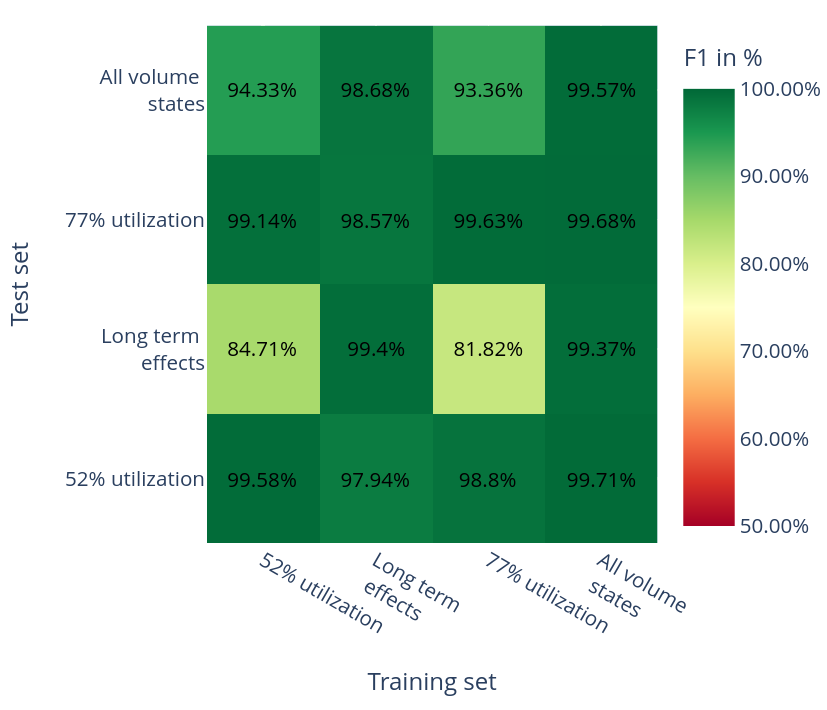} 
    \vspace{-25pt}
	\caption{Results for different volume states (high and medium utilization, long-term effects, and combining all). F1 scores are evaluated across all hold-out test sets using traces from an XFS file system and training XGBoost models.}
	\label{fig:volume_drift_xfs}
\end{figure}


Figure~\ref{fig:volume_drift_xfs} shows that models trained on XFS traces are highly sensitive to the volume state. We evaluate the models trained on \SI{52}{\percent} device utilization on traces from simulated aged devices, and observe a large decrease in F1 score of \SI{15.0}{\percent} (\SI{26.15}{\percent} FNR and \SI{0.5}{\percent} FPR). Interestingly, models trained on either EXT4 or NTFS show small performance  decrease in F1 scores.
We study the drop in performance through feature importance analysis. XFS models trained on \SI{52}{\percent} utilization evaluated on test traces from the same volume state show a predominance of throughput, entropy, and LBA-related features. Similarly, we evaluate  the same model on the long-term data placement setup. The feature analysis shows an increase of importance of entropy features and a negative contribution of certain LBA derived features. We determine that LBA-related features are valuable in a medium disk utilization setup. Nevertheless, they are unreliable for aged devices; there, fragmentation has severely changed the LBA profile. Similarly, the features are also unreliable when using a model on devices with high disk utilization or with significantly different storage capacities. Therefore, LBA features can enhance performance only when evaluating the models on setups similar to those used for training.

We observe that models trained on traces from a single file-system type tend to share similiar feature importance even when trained on different volume states. In particular, EXT4-based models strongly favor entropy-related features. Specifically, entropy histogram lead the feature importance. These models show more resilience against changes in volume state, which mainly impact LBA-related features. In fact, models trained on EXT4 traces show small performance decreases when evaluated on other volume states. The lowest F1 score is \SI{96.23}{\percent}, and corresponds to models trained on \SI{77}{\percent} utilization and evaluated on traces from the long-term data placement setup. 
Interestingly, NTFS-based models rely heavily on transfer-size related features: There is a large drop in importance value between these features and the remaining ones. This characteristic makes them also resilient to volume state changes. Models trained on NTFS trace show negligible drop in F1 scores when evaluating them across different volume states. In fact, all F1 scores are above \SI{99}{\percent}.

We further analyze the effects from more realistic aging effects by validating our model with all volume states using Geriatrix with the Agrawal profile~\cite{geriatrix} and observe improved results confirming the robustness of the model: \SI{97.78}{\percent} (stddev \SI{0.73}) and \SI{96.92}{\percent} (stddev \SI{0.89}) F1-scores for \SI{1000} disk overwrites, and \SI{96.91}{\percent} (stddev \SI{0.77}) and \SI{96.78}{\percent} (stddev \SI{0.73}) for \SI{10}{\kilo{}} overwrites.

Finally, we perform a cross-evaluation on the same set of traces using our proposed features and compare them against the features used in~\cite{hirano2019machine}.
The evaluation on XFS, EXT4, and NTFS traces show a drop in median F1 score using the latter features of \SI{1.67}{\percent}, \SI{6.14}{\percent}, and \SI{10.05}{\percent} respectively, as well as minimal F1 scores of \SI{87.5}{\percent}, \SI{89.56}{\percent}, and \SI{87.72}{\percent} for each file system type. This non-negligible decrease in performance shows the importance of including a larger and more expressive set of features as small changes in F1 scores can lead to large false negative rates or unmanageable FPR. 

\vspace{-5pt}

\subsection{File-system generalizability}
\label{filesystem_generalizability}

Different applications require specific OS-based functionalities making robust and general models necessary.
To study the generalizability of ML models across file-system types we combine traces from all volume states. Although training filesystem-specific models is a possibility, maintaining several ML models in production is a challenge due to the increased memory consumption. This can be mitigated by using a single robust model and using the file-system type as a feature. 
The training and test split of the traces was performed using the same criterion as for the previous analysis. 

\begin{figure}[!b]
    \vspace{-10pt}
	\centering
	\includegraphics[width=\linewidth]{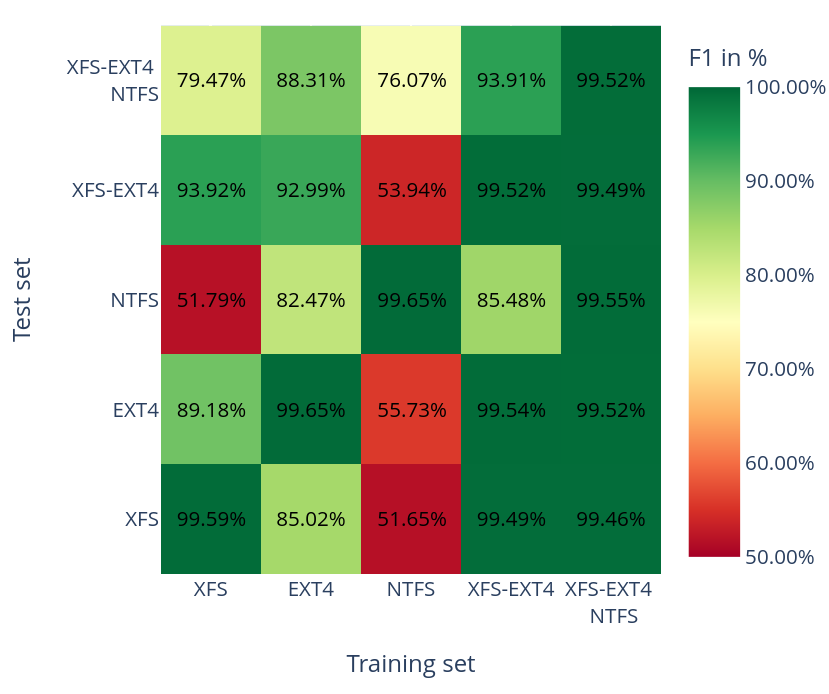} 
    \vspace{-20pt}
	\caption{Generalizability across file system types. F1 scores of XGBoost models trained and evaluated on either XFS, EXT4, and NTFS.}
	\label{fig:f1_filesystem}
\end{figure}

Figure \ref{fig:f1_filesystem} shows a large performance drop when training on XFS and evaluating of NTFS traces (F1: \SI{51.79}{\percent}, FPR: \SI{3.01}{\percent}, FNR: \SI{64.01}{\percent}) and when training on NTFS and evaluating on Linux file systems, namely XFS (F1: \SI{51.65}{\percent}, FPR: \SI{14.12}{\percent}, FNR: \SI{60.26}{\percent}) and EXT4 (F1: \SI{55.73}{\percent}, FPR: \SI{9.0}{\percent}, FNR: \SI{57.89}{\percent}). 
Interestingly, models trained on EXT4 traces have the lowest FNR when evaluated on a different file system type; FNR being \SI{7.79}{\percent} on XFS and \SI{4.2}{\percent} on NTFS. Nevertheless, they are biased toward higher FPR rates: \SI{24.69}{\percent} on XFS and \SI{36.55}{\percent} on NTFS. Despite lower F1 scores, XFS models have better FPR of \SI{1.72}{\percent} on EXT4 and \SI{3.01}{\percent} on NTFS. When evaluating XFS or NTFS-derived models on a different file system, there is a small FPR increase but a disproportionately larger FNR with some values above \SI{50}{\percent}. 

We perform feature permutation analysis on the XFS model and evaluate on NTFS traces; the analysis reveals increased importance of certain entropy features and negative contributions from transfer-size derived features. These negative contributions originate from transfer size differences between Linux and Windows file systems. On the other hand, models trained and evaluated on NTFS traces reveal a heavy reliance on transfer size and very small contributions from remaining features. This leads to low performance when evaluating on Linux filesystems. 
In fact, we observe strong negative contributions from transfer size features and read-LBA features. This seems to confirm our previous hypothesis for performance decrease which originates from transfer size-specific and LBA-related features; these features become meaningless when changing either the volume state or the file system type. 
Furthermore, including file-system awareness helps to improve the performance when training on all file system types by up to \SI{2}{\percent}. 
An evaluation of our traces using features from \cite{hirano2019machine} results in a higher median FPR of \SI{6.2}{\percent} compared to \SI{0.3}{\percent} using our features as well as lower F1 of \SI{93.08}{\percent} compared to \SI{99.47}{\percent}. Therefore, our features lead to more robust models when performing inference across multiple filesystems. This is particularly relevant for data centers dealing with wide range of workloads and server configurations using potentially different file systems.

\vspace{-5pt}
\subsection{Benign workloads generalizability} \label{workloads_generalizability}

For this analysis, we aim to assess the FPR of our model on multiple unseen benign workloads. To do so, we create training data sets by combining traces from all volume states and  use emulated ransomware, file conversion, and compression traces (using \texttt{ZIP} and \texttt{LZMA} algorithms). In addition, we either add OLTP workloads using \texttt{MySQL} with built-in compression disabled (denoted as MySQL w/o CMP), OLTP workloads using \texttt{PostgreSQL} (denoted as PGSQL) and a combination of both sets. The OLTP workloads generated by \texttt{sysbench} use a Pareto random number distribution. Changing the random number distribution is useful for simulating OLTP workloads resulting from different types of transaction data. 

The test sets we use to evaluate our model consist of file compression using \texttt{BZ2} algorithm, \texttt{MySQL} traces without compression obtained using a uniform random number distribution for the transactions, \texttt{MySQL} traces using Pareto distribution with built-in compression, \texttt{MySQL} traces from a uniform distribution with built-in compression, and finally \texttt{PostgreSQL} traces using a uniform distribution.

\begin{figure}[t]
	\centering
	\includegraphics[width=\linewidth]{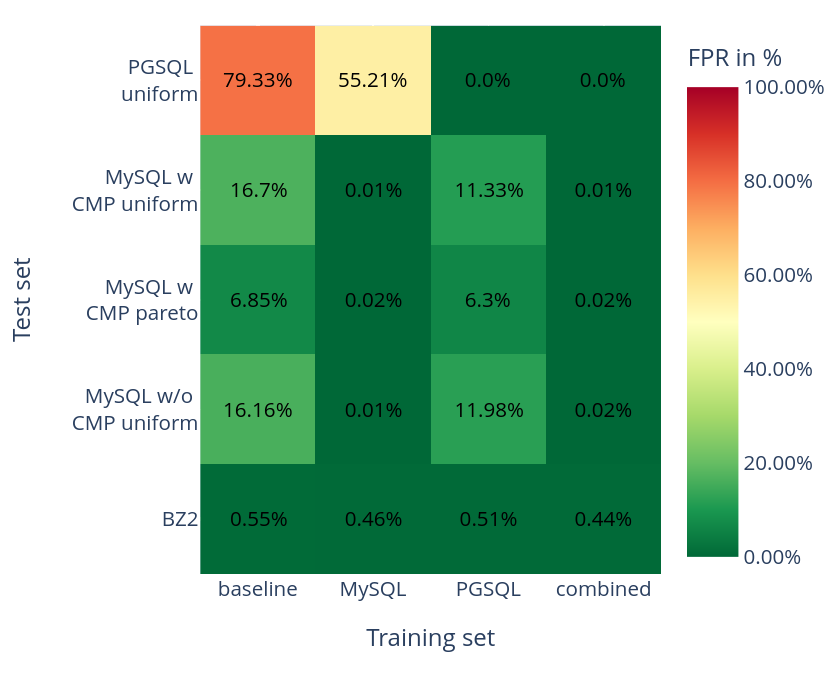} 
        \vspace{-25pt}
	\caption{FPR of benign workloads trained on either baseline,  MySQL, PGSQL or
        combined with XFS using our features.}
        \vspace{-10pt}
	\label{fig:fpr_workloads_XFS}
\end{figure}
Figure \ref{fig:fpr_workloads_XFS} shows that the model trained on the baseline dataset leads to low FPR when evaluated on traces from \texttt{BZ2} compression. As we train on  compression traces, we expect that similar entropy values and overall behavior lead to high accuracy. On the other hand, evaluating this model on OLTP traces leads to high FPR.
We perform feature importance of this model on \texttt{PostgreSQL} traces, and observe negative contributions of several LBA-related features. Including a subset of \texttt{MySQL} traces (second column) reserved for training significantly reduces the FPR to less than \SI{1}{\percent} for remaining \texttt{MySQL} traces and does not increase the FPR of \texttt{BZ2} traces. Including benign workloads with different IO patterns has the most significant impact in reducing FPR and is critical for efficent ransomware detection. 
In spite of the increased performance on \texttt{MySQL} traces, \texttt{PosgreSQL} remains challenging to classify, with an FPR of \SI{30.89}{\percent}  on EXT4 and \SI{55.21}{\percent} on XFS. Feature importance analysis shows several negative contributions led by read throughput and LBA-related features. OLTP traces are characterized by medium entropy values and a heavy number of writes. 

To have a fine-grained comparison between ransomware and the database drivers used, we analyze the distributions for the average 4K entropy, write and read throughput. \texttt{PostgreSQL} has a higher number of reads, which may lead to models that are heavily reliant on read throughput for detecting ransomware. XFS traces have a median FPR of \SI{0.5}{\percent} compared to \SI{0.4}{\percent} for EXT4. Nevertheless, the lowest performance leads to \SI{79.33}{\percent} FPR compared to \SI{46.47}{\percent} for EXT4; this corresponds to training the model on the baseline dataset and evaluating it on \texttt{PosgreSQL} traces. 

\begin{figure}[!t]
	\centering
	\includegraphics[width=\linewidth]{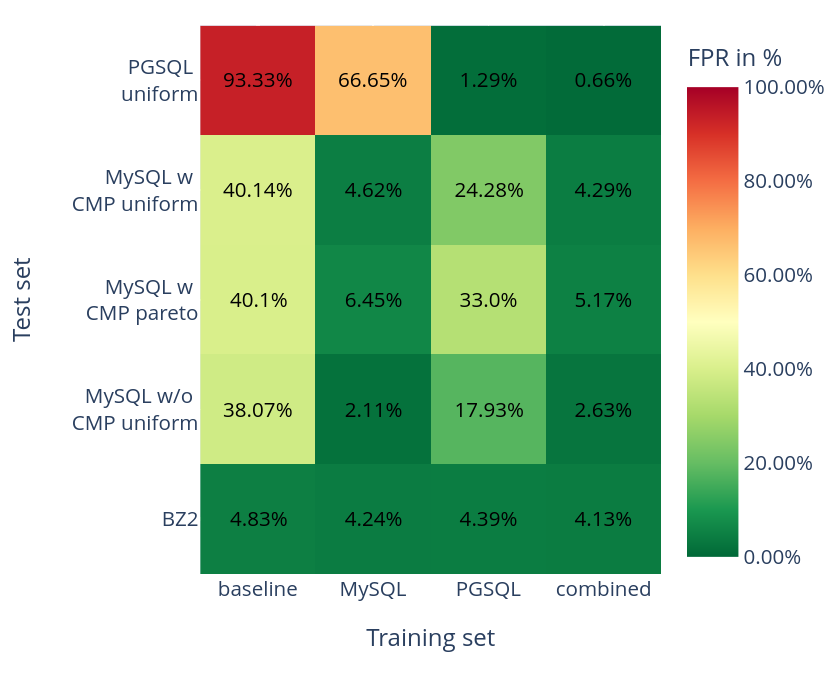} 
         \vspace{-25pt}
	\caption{FPR of benign workloads trained with XFS on either baseline,  MySQL, PGSQL, and combined traces using features from \cite{hirano2019machine}.}
	\label{fig:fpr_workloads_XFS_hirano}
          \vspace{-10pt}

\end{figure}

Next, we evaluate our EXT4 model trained using the features from Hirano et al.~\cite{hirano2019machine} (Figure ~\ref{fig:fpr_workloads_XFS_hirano}). We obtain a higher median FPR of \SI{5}{\percent} and \SI{9.7}{\percent} for EXT4. Similarly, we train XFS models on baseline and \texttt{MySQL} datasets and evaluate them on \texttt{PosgreSQL} traces; these features lead to an increase of \SI{14}{\percent} and \SI{14.44}{\percent} respectively in FPR.
\begin{figure}[!b]
	\centering
	\includegraphics[width=0.9\linewidth]{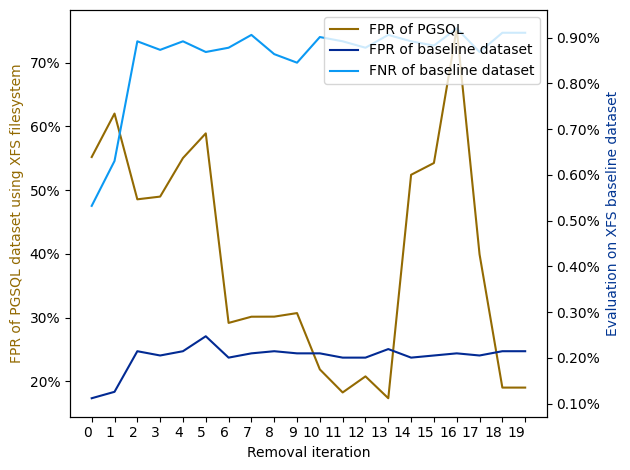} 
	\caption{Greedy feature removal of XGBoost model trained on \texttt{MySQL} dataset using XFS file system.}
	\label{fig:XFS_evolution_nofeatures}

\end{figure}

Additionally, we study the impact of successively and greedily removing features from our model (Figure~\ref{fig:XFS_evolution_nofeatures}). The process runs 20 iterations, during each one the permutation analysis is computed and the lowest-performing feature is removed. Although this strategy is unable to foresee combinations of features that might lead to the lowest FPR possible, it is a first evaluation to determine if any features could be dropped without harming or potentially even improving the accuracy. We plot the evolution of the FPR computed on the PostgreSQL dataset over the removal iterations to determine the minimal number of features to be removed leading to the largest decrease in FPR. Furthermore, we plot both the FPR and FNR on our baseline dataset to assess the impact on baseline performance. We observe that the removal of the 12 least important features reduces the FPR from \SI{55}{\percent} to less than \SI{20}{\percent} and does not severely impact either the FPR or FNR on the baseline dataset. After removing the 13th feature, the FPR rapidly rises indicating the minimum at 12 removed features.

\vspace{-5pt}
\section{Copy-on-write effects from VM images} \label{QCOW_effects}

VM images use formats such as \texttt{qcow2} that store data as a single file. Data modification within the VM results in very different LBA access patterns and even write entropy compared to modifying the data directly on the device. Therefore, it is relevant to assess the generalizability of our approach by training models on activity from either XFS or EXT4 and evaluating these on traces from VM environments. 

To do so, we first instantiate a Windows 10 VM using NTFS stored as a \texttt{qcow2} image on top of either an XFS or EXT4 device on the host. For this analysis, we include file conversion, compression, OLTP simulations, and ransomware traces from all volume states on XFS or EXT4. To create disjoint training and test datasets, we assign traces based on thread concurrency. Inspection of IO patterns reveals a noisier entropy distribution for both ransomware and benign samples compared to our previous XFS and EXT4 traces. To evaluate our model's performance we create 5 datasets: EXT4, \texttt{qcow2} traces on top of EXT4 (e.g., EXT4QCOW2), aggregation of EXT4 and NTFS, aggregation of EXT4 and EXT4QCOW2, and finally a dataset containing all 3 different datasets. We include traces from NTFS for certain datasets as EXT4QCOW2 traces were obtained from a device using EXT4 but the Windows VM uses NTFS which might lead to models trained on EXT4 and NTFS traces performing better in this case. 

\begin{figure}[!b]
	\centering
	\includegraphics[width=\linewidth]{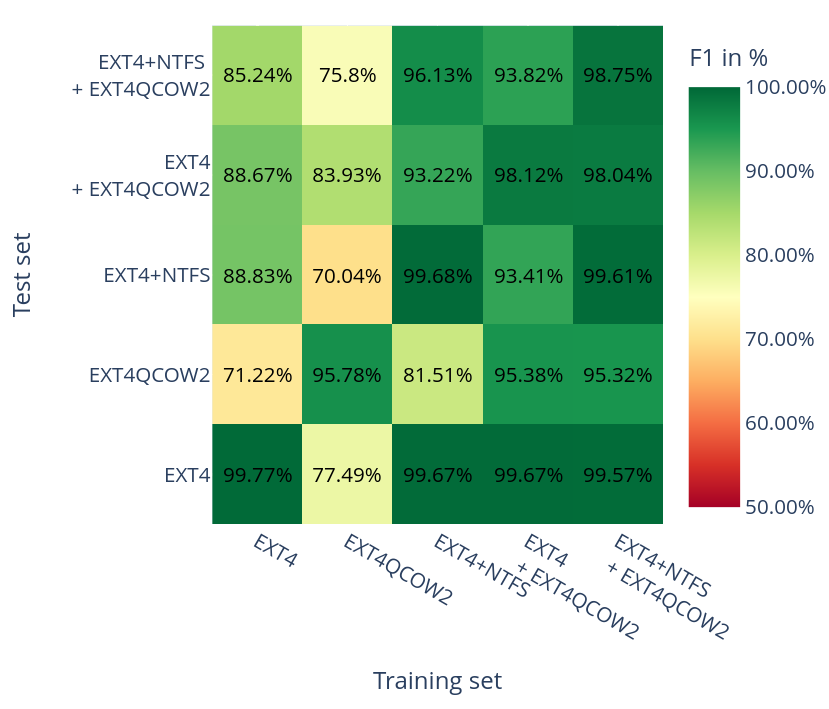} 
        \vspace{-20pt}
	\caption{F1 scores of XGBoost models trained and evaluated on a combinations EXT4, NTFS and \texttt{qcow2} traces on top of an EXT4 device (EXT4QCOW2) without VM encryption.}
	\label{fig:f1_ext4qcow_systor}
\end{figure}

Figure~\ref{fig:f1_ext4qcow_systor} shows that models trained only on EXT4 traces exhibit poor performance on EXT4QCOW2 traces (F1: \SI{71.22}{\percent}, FPR: \SI{43.63}{\percent}, FNR: \SI{20.57}{\percent}). 
Although training on EXT4QCOW2 traces leads to a good F1 score when evaluating on the same setup, the value is lower compared to training and evaluating on EXT4. In fact, obtaining the ground truth is harder as for \texttt{qcow2} traces writes might not only correspond to active ransomware but also to OS activity. This leads to a noisy entropy distribution which indicates IOs from ransomware and OS activity happening simultaneously. 
Interestingly, evaluating an EXT4QCOW2 model on EXT4+NTFS traces results in lower accuracy than evaluating only on EXT4. Permutation importance on this model shows a predominance of entropy and throughput-related features and negative contributions from LBA features. In fact, data allocation on a \texttt{qcow2} image as seen from the LBA sector on the device differs greatly from NTFS or EXT4 traces. 
Training on both EXT4 and NTFS traces and evaluating on EXT4QCOW2 leads to a \SI{10.29}{\percent} increase in F1 score compared to training only on EXT4. Feature importance comparison between the two training sets shows differences of importance for the read throughput which is the most important feature when training on EXT4+NTFS and 4th most important when training only on EXT4. Finally, including both EXT4 and EXT4QCOW2 traces leads to an overall good accuracy on all testing sets. Training on both EXT4 and EXT4QCOW2 traces resulting from VM activity is necessary to ensure robustness when deploying ransomware detection in virtualized environments. Evaluation of this model on EXT4+NTFS traces shows good accuracy with \SI{0.93}{\percent} FPR and \SI{11.55}{\percent} FNR. 

\begin{figure}[!b]
	\centering
	\includegraphics[width=\linewidth]{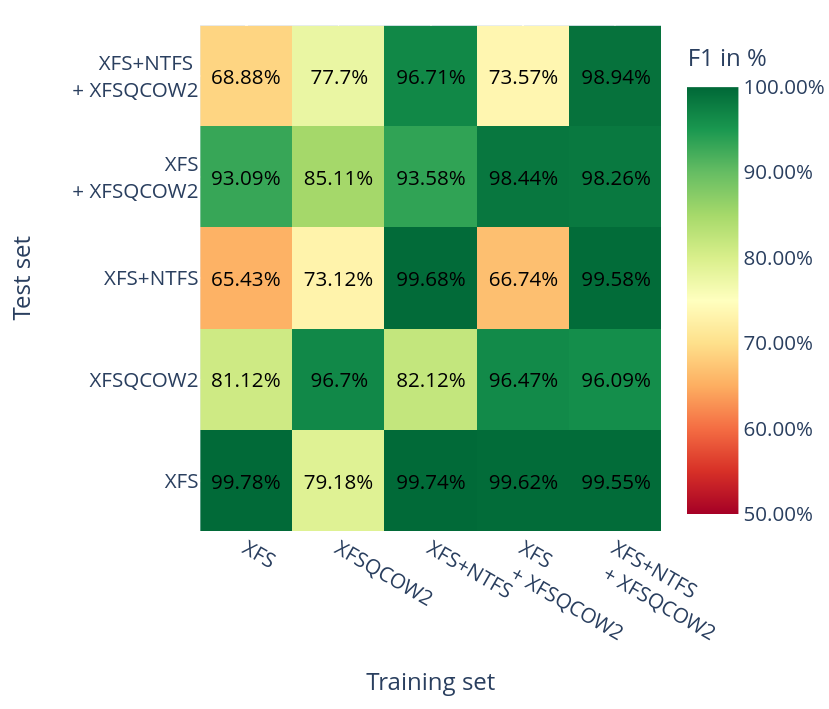} 
     \vspace{-20pt}
	\caption{F1 scores of XGBoost models trained and evaluated on a combination XFS, NTFS, and \texttt{qcow2} traces on top of an XFS device (XFSQCOW2) without VM encryption.}
	\label{fig:f1_xfsqcow_systor}
\end{figure}

We perform a similar cross-evaluation using XFS-based traces instead of EXT4 with identical workloads in Figure~\ref{fig:f1_xfsqcow_systor}. XGBoost models trained on XFS evaluated on XFSQCOW traces show a higher F1 of \SI{81.12}{\percent} compared to EXT4. Similar to EXT4 results, XFSQCOW2 models evaluated on XFS traces show a moderate F1 score of \SI{79.18}{\percent} as well as XFS+NTFS models evaluated on XFSQCOW traces an F1 score of \SI{82.12}{\percent}. Nevertheless, XFS and XFSQCOW2 models evaluated on the XFS+NTFS model show a steep drop in F1 score. As shown in our file-system generalizability analysis, models trained on XFS traces perform poorly on NTFS traces which explains the \SI{65.43}{\percent} F1 score.

\section{Device encryption} \label{device_encrytion}

In this section we study device encryption using LUKS and VMs protected with Bitlocker encryption.
\vspace{-10pt}
\subsection{LUKS encryption} \label{LUKS_encrytion}

LUKS introduces an additional device mapper layer that affects LBA access patterns and entropy of writes. For this analysis, we use file conversions, compression, and OLTP workloads as benign traces and assign them to training or test sets based on the concurrency. 

\begin{figure}[!b]
	\centering
    \vspace{-10pt} 
	\includegraphics[width=\linewidth]{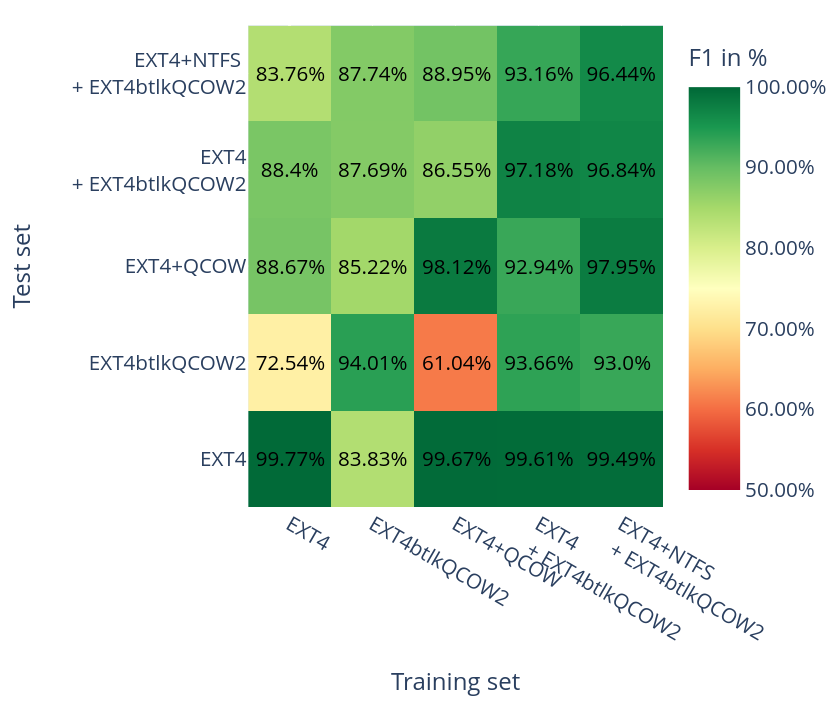} 
    \vspace{-25pt}
	\caption{F1 score of XGBoost models trained and evaluated on combinations of EXT4, EXT4QCOW2 and traces \texttt{qcow2} VM with Bitlocker encryption on top of an EXT4 device (EXT4btlkQCOW2).}
	\label{fig:f1_ext4bqcow_systor}
\end{figure}

XGBoost trained on EXT4 traces without encryption is unable to maintain high accuracy when evaluated on EXT4 with LUKS (FPR of \SI{59.11}{\percent}, and FNR of \SI{29.79}{\percent}). On the other hand, models trained on EXT4 with LUKS can perform well on EXT4 traces with an FPR of \SI{10.05}{\percent} and an FNR of \SI{7.16}{\percent}. Feature importance analysis of models trained on EXT4 traces evaluated on EXT4 with LUKS encryption shows a predominance of read throughput and entropy-related features as well as a negative contribution from transfer size and LBA-derived ones. The importance of the MAD of the write transfer size is reduced by \SI{47.9}{\percent} when using encryption. 
Detection based only on average entropy is no longer viable with encrypted devices as both ransomware and benign workloads result in high entropy write IOs. 
Our evaluation using the features from~\cite{hirano2019machine} of an EXT4 trained model on EXT4 with LUKS encryption results in an \SI{11.99}{\percent} decrease in F1 score as well an increase of \SI{27.11}{\percent} of the FPR and \SI{16.37}{\percent} of the FNR. 

The situation is yet different for XFS: When training on XFS without encryption and evaluating on XFS with LUKS, the F1 score is \SI{37.94}{\percent} higher than the previous evaluation using EXT4-based traces. Nevertheless, when training XFS with LUKS, the model suffers from extremely poor performance with an unacceptable F1 score of \SI{1.14}{\percent}, and an FPR of \SI{0.01}{\percent} and an FNR of \SI{99.43}{\percent}. 
Feature importance analysis of the latter model shows a predominance of LBA and transfer size-related features as entropy-based detection is less viable when encrypting the device. Evaluation using the features from~\cite{hirano2019machine} of an XFS trained model evaluated on XFS with LUKS encryption results in a \SI{19.94}{\percent} decrease in F1 score, an FPR of \SI{2.04}{\percent} but an FNR increase of \SI{20.3}{\percent}, although training on XFS-LUKS and testing on XFS results in a \SI{0.8}{\percent} increase in F1 when using the features from~\cite{hirano2019machine} compared to ours.
\vspace{-5pt}

\subsection{Encrypted VM environments} \label{QCOW_bltk}

Similarly to LUKS encryption in Linux, Windows QEMU/KVM VMs support Bitlocker encryption. We expect enabling this feature to result in hard-to-classify traces as entropy-based detection is less viable and the traces include noise from the OS activity. For this analysis, we included similar benign and ransomware workloads as in the previous sections. To evaluate changes in our model's accuracy we designed again 5 datasets: EXT4, \texttt{qcow2} VM with Bitlocker encryption on top of EXT4 (EXT4btlkQCOW2), combination of EXT4 and EXT4QCOW2, combination of EXT4 and EXT4btlkQCOW2, and finally a dataset containing all 3 different datasets. In this analysis, we replace NTFS traces with EXT4QCOW2 traces because we want to test whether including EXT4QCOW2 helps the model to classify BitLocker-encrypted \texttt{qcow2} traces. Despite enabling VM encryption, IO inspection of EXT4btlkQCOW2 shows slight differences in mean entropy values between ransomware and benign traces. Further analysis of mean entropy histograms of EXT4QCOW2 traces with benign and ransomware stratification using 20 bins shows broad distributions ranging from low to maximal entropy values, with the ransomware's distribution centered around high entropy values. On the other hand, EXT4btlkQCOW2 has very narrow distributions with two (ransomware) or three modes (benign). For ransomware traces almost all values lay on the highest entropy bin (\SI{87}{\percent}), whereas benign traces have a significant proportion of values on the lowest entropy bin (\SI{25}{\percent}).

The comparison of XGBoost models trained on EXT4 and evaluated on EXT4btlkQCOW2 have a \SI{1.73}{\percent} increase in F1 score w.r.t. evaluation on EXT4QCOW2 (Figure~\ref{fig:f1_ext4bqcow_systor}). Furthermore, models trained on EXT4btlkQCOW2 show good performance on all setups, with even higher values when evaluating on EXT4 traces (\SI{83.83}{\percent}) compared when training on EXT4QCOW2 (\SI{77.49}{\percent}). Including EXT4 and EXT4QCOW2 traces in the training harms the performance producing an \SI{11.5}{\percent} decrease in the F1 score. On the other hand, including EXT4 and EXT4btlkQCOW2 leads to high-performing models on all setups, EXT4QCOW2 included.

\begin{figure}[t]
	\centering
	\includegraphics[width=\linewidth]{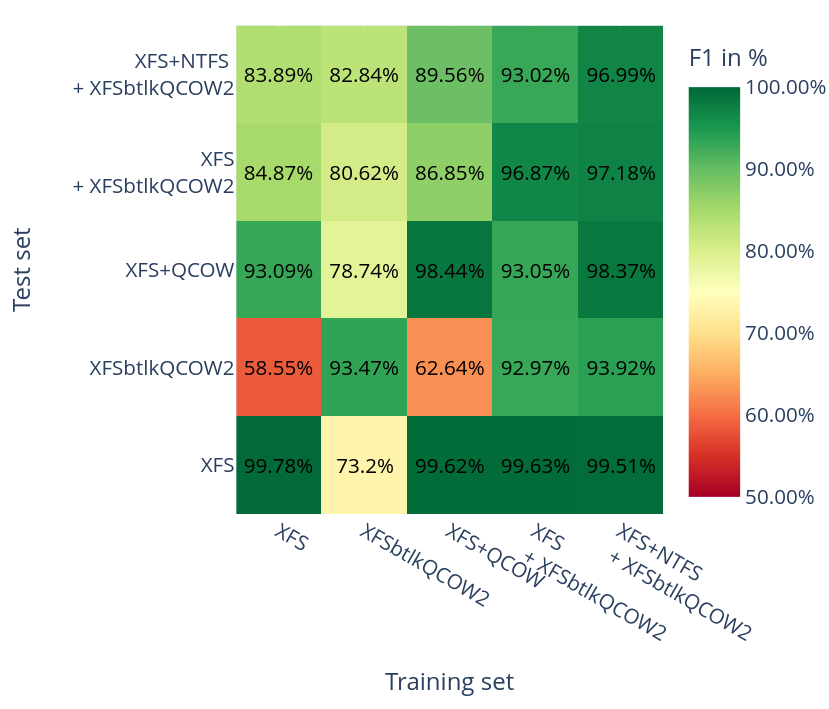} 
     \vspace{-25pt}
	\caption{F1 score of XGBoost models trained and evaluated on combinations XFS, XFSQCOW2 and traces \texttt{qcow2} VM with Bitlocker encryption on top of an XFS device (XFSbtlkQCOW2).}
	\label{fig:f1_xfsbqcow_systor}
    \vspace{-15pt}
\end{figure}

We elaborate an equivalent dataset using XFS instead of EXT4 and perform similar analyses. As Figure~\ref{fig:f1_xfsbqcow_systor} shows, models trained on XFS and evaluated on XFSbtlkQCOW2 lead to a \SI{22.57}{\percent} drop in F1 score with encryption when compared to evaluating on XFSQCOW2. Training an XGBoost model on XFSbtlkQCOW2 traces leads to a modestly performing model across all setups, albeit with lower F1 scores than the symmetric EXT4 cross-evaluation. Using this file system, adding XFSQCOW2 on top of XFS traces increases the F1 score by \SI{3.69}{\percent}. Finally, training a model on XFS and XFSbtlkQCOW2 leads to high accuracy on all setups.

Overall, existing models do not generalize to encrypted workloads. However, by adding training data from encrypted workloads, a model can learn to detect ransomware efficiently indicating the importance of non-entropy based features. Care must be taken when deploying encrypted VM images as training data from simple block-level encryption does not generalize to these workloads, hence requiring to augment the training set with such traces.
Finally, our feature sets provides more robust models than others.

\section{Real-world setup}
\label{sec:real}
\vspace{-5pt}

When using our storage system with CSDs, we were not able to measure any throughput and latency degradations. Therefore, we focus on architecture a from Figure~\ref{fig:arch_overview} to evaluate the overhead when hardware-based feature extraction using CSDs is not available.

\subsection{Inline ransomware detection}

We use \texttt{sysbench} with varying number of threads to simulate OLTP workloads representative of real-world activity. The \texttt{sysbench} built-in functionalities allow us to assess both latency and throughput effects, for both read and write operations, introduced by our \texttt{dm-entropy} kernel module as well as \texttt{SystemTap}. As our baseline, we execute \texttt{sysbench} without our kernel module. We then study the impact of introducing both tools with the entropy calculation disabled, therefore collecting the previously described information except for the entropy. Finally, we enable entropy extraction and studiy the full \texttt{dm-entropy} setup which we then used to verify real-time ransomware detection.

\begin{figure}[t]
	\centering
	\includegraphics[width=\linewidth]{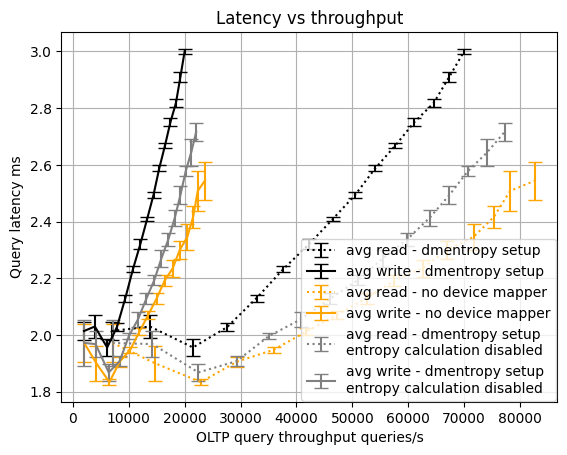} 
	\caption{Latency vs throughput analysis of our \texttt{dm-entropy} setup. We use \texttt{sysbench} to run commonly used \texttt{MySQL} database workloads. Different throughput levels were obtained by varying the thread concurrency levels. A reduction in latency or throughput leads to curves closer to the origin.
 }
	\label{fig:oltp_stress_test}
    \vspace{-10pt}
\end{figure}

Figure~\ref{fig:oltp_stress_test} shows bandwidth reduction and latency increase when using the kernel module to extract IO information. Our pipeline introduces an average latency increase of only \SI{268}{\micro\second} (\SI{12.04}{\percent})  with a standard deviation of \SI{124}{\micro\second} and a low maximal throughput reduction of \SI{15.3}{\percent} for both read and write queries. 
Furthermore, results from our setup with the entropy calculation disabled are relatively close to the setup using no device mapper indicating entropy calculation as the main factor for the increased latency. Extracting entropy directly in hardware using a CSD is an attractive alternative for reducing such effects.

\begin{figure}[t]
	\centering
	\includegraphics[width=\linewidth]{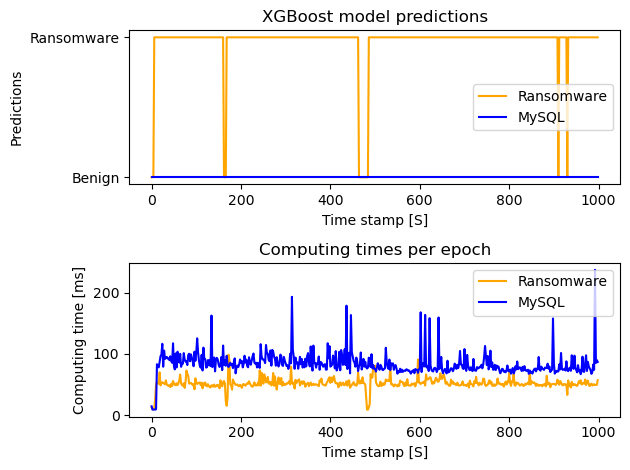} 
    \vspace{-20pt}
	\caption{Real-time ransomware detection using our pipeline. Collecting IO information using \texttt{dm-entropy} and deriving features is computationally efficient and constitutes an interesting alternative to OS-level detection (lower plot).}
	\label{fig:online}
    \vspace{-10pt}
\end{figure}

Furthermore, as Figure~\ref{fig:online} shows, deriving our set of features as well as ML inference can be done within \SI{200}{\milli\second}. 
In contrast, in our storage system with CSDs we measure ML inference time less than \SI{15}{\milli\second} for more than \SI{2000} volumes thanks to the SnapML library~\cite{snapboost}. 
For every epoch, we extract the features and perform inference on the latest IO traces and use \texttt{time.sleep()} on the remaining time of our window. Therefore, given a fixed window for collecting IO traces, e.g. T=\SI{5}{\second}, we detect an attack after at most \SI{5.2}{\second} from the start of the ransomware encryption using the first-level classifier.
Note that additional post processing of a several consecutive epochs in a sliding window manner to further reduce the FPR will prolongate the detection time.
Increasing the throughput will lead to longer computing times as the number of IOs to collect is larger and the extraction of the also takes longer. As we collect and clean the IO traces using bash utilities and extract the features on Python, we expect the computing time to be further reduced when using a pipeline written in C.

\subsection{Real ransomware validation}
\label{real_ransomware_validation}

Finally, we assess our ransomware detection pipeline on real ransomware collected from Malware Baazar~\cite{malwarebazaar}. To prevent any security issues, we run all samples inside a QEMU VM using either Windows 10 or Fedora 40 and with no network interface. Black Basta, Conti, Lockbit, Lockfile, Defray, Sodinokibi, Wannacry~\cite{Rehman}, and Underground are the ransomware samples running on Windows. On Linux we use Conti, Erebus~\cite{Hawawreh}, Monti, Sodinokibi, RansomExx (also known as Defray), and BlackCat. Samples run for at least \SI{1}{\hour} or until the process exited for fast encrypting ransomwares. Furthermore, on Windows we studied ransomware samples running on VM images stored either on raw or QCOW format as copy-on-write has a significant impact on detectability (See Section \ref{QCOW_effects}). We train a model using samples from all previous traces from benign workloads and from the aforementioned WannaLaugh ransomware emulator as well as using \SI{1}{\percent} from collected real ransomware traces. 

As Table~\ref{tab:win_ransomware} shows, our model can accurately detect ransomware in most setups. Lockbit remains challenging to classify as it is characterized by encrypting only the first \SI{4}{\kibi\byte} resulting in faster encryption.  Performance variations between \texttt{raw} and \texttt{XFSqcow2} can be explained by the strong effects of copy-on-write setups on IO block operations. Performance on Linux setups is high for the majority of the ransomware samples as can be seen in Table~\ref{tab:fedora_ransomware}.

\begin{table}[t]
\centering
\begin{tabular}{|c|c|c|c|c|}
\hline
\multirow{3}{*}{Samples} & \multicolumn{4}{|c|}{False negative rates} \\
\cline{2-5}
& \multicolumn{2}{|c|}{raw} & \multicolumn{2}{|c|}{XFSqcow2}\\
\cline{2-5}
& FS1 & FS2 & FS1 & FS2 \\
\hline
Blackbasta & \textbf{0.0} & 1.2 & \textbf{1.0} & 13.0\\
Conti & \textbf{0.5} & 1.1 & \textbf{4.2} & 12.4 \\
Lockbit &  \textbf{19.0} &  60.3  &  \textbf{16.3} &  65.1 \\
Lockfile & \textbf{1.3} & 12.4 & \textbf{4.6} & 9.3\\
Defray & \textbf{0.2} &  2.4 &   \textbf{4.6} &  5.8 \\
Wannacry & \textbf{7.1} & 12.8 & \textbf{7.6} & 18.7 \\
Underground & \textbf{1.8} & 8.0 & \textbf{11.9} & 19.9 \\
Sodinokibi & \textbf{0.0} & 0.3 & \textbf{0.7} & 5.3 \\
\hline
\end{tabular}
\vspace{-0.2cm}
\caption{Inference results on real ransomware traces from Windows 10 VMs in percent. FS1 corresponds to model training using our features and FS2 to the features from~\cite{hirano2019machine}.}
\label{tab:win_ransomware}
\end{table}

\begin{table}[b]
\centering
\begin{tabular}{|c|c|c|c|c|}
\hline
\multirow{3}{*}{Samples} & \multicolumn{4}{|c|}{False negative rates} \\
\cline{2-5}
& \multicolumn{2}{|c|}{Raw XFS} & \multicolumn{2}{|c|}{Raw EXT4} \\
\cline{2-5}
& FS1 & FS2 & FS1 & FS2\\
\hline
Erebus & \textbf{1.8} & 19.8 & \textbf{0.3} & 8.8\\
Conti & \textbf{0.0} & 0.2 & \textbf{0.0} & 0.1  \\
Monti & 4.1 & \textbf{3.7} & \textbf{0.1} & 0.3 \\
RansomExx & \textbf{2.1} & 3.6 & \textbf{2.1} & 18.2 \\
BlackCat & \textbf{1.4} & 5.2 & \textbf{0.0} & 1.4 \\
Sodinokibi & 1.1 & \textbf{1.0} & \textbf{2.7} & 13.2 \\

\hline
\end{tabular}
\vspace{-0.2cm}
\caption{Inference results on ransomware traces from a Fedora VM in percent. FS1 corresponds to model training using our features and FS2 to \cite{hirano2019machine} features.}
\label{tab:fedora_ransomware}
\end{table}


\section{Related Work}
\label{sec:related}




Shinagawa et al.~\cite{Shinagawa2009} developed a new hypervisor architecture called BitVisor, allowing most IO access from the guest OS to pass through; the hypervisor would only handle a small number of necessary IOs to maintain security functionalities. This hypervisor architecture paved the way for Hirano et al.~\cite{hirano2017} to build WaybackVisor to collect IO logs by saving them to a Hadoop Cluster. This setup allowed them to collect traces from benign workloads and ransomware attacks. 
They collected activity from WannaCry and TeslaCrypt ransomware, as well as Zip file compression to simulate benign workloads.
Their pipeline reached F1 scores as high as \SI{98}{\percent} for ransomware and wiper-malware classification on NTFS using \SI{10}{\second} windows. 
Despite achieving high accuracy, their analysis was limited to a specific setup and effects from different file systems, file-system aging, or more realistic workloads remain unknown. Using our proposed pipeline, compared to the Hypervisor-Hadoop cluster log collection, we observe significant advantages from efficiently extracting computationally light features from minimal IO information at run time, limiting the trace collection to the necessary data for feature computation and enabling real-time detection at low overhead.

More recently, Zhongyu et. al. \cite{zhongyu} developed \texttt{DeftPunk}, a ransomware detection and recovery framework for block storage in the cloud. Their pipeline collects IO operations using proxies and computes simple features using a sliding window of \SI{10}{\second}. These features allow a first level decision-tree classifier to perform fast scans with a low computational budget. A second level XGBoost classifier is used only on samples previously labelled as ransomware. Their pipeline incorporates a snapshot functionality allowing them to restore the victim's data in case an attack is detected. Nevertheless, the study is limited to append-only distributed filesystems used in cloud environments. In a wide range of workloads and setups the accuracy cannot be externally verified in their cloud-based environment due to the lack of an interface to access the model's output. Table \ref{tab:ransomware_research_overview} shows an overview of the studies on variations of different setups dimensions, showcasing the need for a comprehensive study. 

Using tools such as Wannalaugh~\cite{diamantopolous2024wannalaugh} allow researchers to study a wide range of ransomware and potential evasion techniques by varying the emulator's parameters. Further research is still required to assess an adversarial setting were ransomware can dynamically modify its behavior to evade detection, potentially by leveraging the ML models themselves if exposed during an attack. Additional research is required to quantify possible training data poisoning when working on distributed systems for gathering ML data as performed by Severi et. al~\cite{severi2023poisoning} for malicious network deep-learning classifiers.

Li et al.~\cite{li2018femu} developed FEMU, a QEMU-based flash emulator to replace the SSD. This open-source software can preserve fairly good accuracy with latency variance around \SI{0.5}{} to \SI{38}{\percent}. Modifications of the QEMU stack achieve low latency using up to 32 IO threads enabling large-scale SSD research. 
Our pipeline could be integrated into FEMU thanks to its extensible characteristics.
Firmware-level detection and recovery can provide useful solutions for protecting systems as well as allowing for an efficient restoration of user data~\cite{flashguard, amoeba, ssd_insider}.
ShieldFS~\cite{shieldfs}, a self-healing Windows-based file-system, allows the restitution of the encrypted files when malicious activity is detected. These two approaches could be combined with our ransomware detection to increase the robustness to cyber-threats.

\begin{table}[t]
\centering
\caption{Overview of current research on ransomware detection in storage.}
\vspace{-0.2cm}
\begin{tabular}{|c|c|c|c|}
\hline
\textbf{Studies} & \textbf{Hirano} & \textbf{Zhongyu} & \textbf{Ours} \\ \hline
Ransomware behavior & \cmark & \cmark  & \cmark \\ \hline
Benign workloads & \xmark  & \cmark  &  \cmark \\ \hline
Filesystems & \xmark  & \xmark  & \cmark \\ \hline
Volume state & \xmark  & \xmark  & \cmark \\ \hline
Copy-on-write effects & \xmark  & \cmark  & \cmark \\ \hline
Device encryption & \xmark  & \xmark  & \cmark \\ \hline
\end{tabular}
\label{tab:ransomware_research_overview}
\end{table}

\par
A large part of ransomware research currently focuses on behavioral fingerprinting or Network analysis at the OS-level. Berrueta et al.~\cite{berrueta2019survey} studied state-of-the-art ransomware detection techniques and classified information for ransomware detection into static, dynamic, or network-based data. Both dynamic and network data are obtained when the ransomware is running. Network-based detection is not an uncommon technique as several strains require Internet access to retrieve or store encryption keys or send local data for exfiltration purposes.
Alhawi et al.~\cite{Alhawi_Baldwin_Dehghantanha_2018} developed a pipeline for detecting ransomware attacks on Windows by analyzing network traffic features derived on protocol type as well as duration, packets, and bytes per conversation. On the other hand, dynamic detection encompasses a larger number of patterns such as changes in entropy, monitoring file extensions, honey-pot files, and frequency of API calls as well as system or cryptography library functions~\cite{Alzahrani_Alshahrani_Alshehri_Fu_2019}. Finally, static-based detection aims at detecting malware before it runs by analyzing text strings, function calls, or cryptographic primitives present in the binary~\cite{Zheng_Sun_Lui_2013,Venugopal_Hu_2008, Lestringant} to prevent the execution of malware. Anomaly detection pipelines can effectively detect a large range of malware activity~\cite{Chandola,tang}.  Data-centric approaches, such as CryptoDrop~\cite{cryptodrop} or UNVEIL~\cite{Kharaz}, leverage behavioral information by monitoring transformations of the user data. ML-based detection on behavioral features allow us to approximate complex decision boundaries that lie on high dimensional spaces for which a handmade decision threshold is difficult to design or even intractable~\cite{Maiorca2017, Alzahrani_Alshahrani_Alshehri_Fu_2019}.  

Cohen et al.~\cite{Cohen_Nissim_2018} studied ransomware detection on VMs by analyzing volatile memory dumps and using Random Forest models reaching high accuracy (F1 score \SI{97.6}{\percent}) showing the value of dynamic ML-based detection. 
Ajay et al.~\cite{Ajay_Kumara_Jaidhar_2017} used VM introspection to retrieve semantic details of processes on the guest VM to detect malware activity. Increases in computational power enable deep learning-based detection as an interesting alternative to decision tree-based models, potentially leading to a more expressive being able to analyze fine-grained details as well as harvest the temporal information~\cite{Zahoora_2022, Maniath_Ashok_Poornachandran_Sujadevi_Sankar, Ciaramella_Iadarola_Martinelli_Mercaldo_Santone_2023}. Despite the large body of research on ransomware detection, complementary methods to OS-level defenses are needed to ensure robustness against these attacks.


\section{Conclusions}
\label{sec:conclusion}
Understanding the complex landscape of storage-based ransomware detection requires a deep dive into the high-dimensional space of disk utilization states, file systems, and IO patterns. Our research has demonstrated that while changes in volume state can affect performance across different file systems, comprehensive training across a spectrum of disk utilizations enhances the model's resilience, ensuring sustained accuracy even in intermediate volume states. Moreover, incorporating file-system specific training enhances performance across a diverse array of scenarios, affirming the value of file-system awareness in our models.

The impact of significant file-system level changes, such as the integration of LUKS device encryption or the adoption of copy-on-write VMs, cannot be overstated. As detailed in Sections~\ref{QCOW_effects} and~\ref{device_encrytion}, these alterations profoundly affect IO patterns, necessitating the inclusion of specialized workloads in our training regimen. This tailored approach enabled our models to achieve promising results in detecting real ransomware across Windows and Linux environments, with low false negative rates in most instances. Despite these advances, the ransomware Defray presents unique challenges, indicating the need for further refinement in our models to address such elusive threats.

Our analysis thus far has laid a solid foundation, yet the path forward is ripe with opportunity. Future work will delve into the nuances of emerging ransomware threats, such as very fine-grained intermittent encryption, wiperware, and exfiltration. Additionally, exploring alternative machine learning architectures that leverage the temporal dynamics of IO traces holds the potential to elevate our F1 scores even further. With these avenues for advancement, we remain optimistic about our continued progress in the realm of ransomware detection, poised to explore and innovate in the face of cybersecurity's ever-evolving challenges.
\bibliographystyle{plain}
\bibliography{references}

\end{document}